\documentclass[10pt]{iopart}
\pdfoutput=1
\usepackage{iopams}  

\expandafter\let\csname equation*\endcsname\relax
  \expandafter\let\csname endequation*\endcsname\relax

\usepackage{amssymb}
\usepackage{amsmath}

\usepackage{soul}
\usepackage{cite}
\usepackage{enumerate}

\usepackage{empheq}

\usepackage{fancybox}

\usepackage{etoolbox}
\usepackage{hyperref}

\usepackage{color}

\makeatletter
\def\@mkboth#1#2{}
\newlength\appendixwidth
\preto\appendix{\addtocontents{toc}{\protect\patchl@section}}
\newcommand{\patchl@section}{%
  \settowidth{\appendixwidth}{\textbf{Appendix }}%
  \addtolength{\appendixwidth}{1.5em}%
  \patchcmd{\l@section}{1.5em}{\appendixwidth}{}{\ddt}%
}
\makeatother

\begin{document}


\title[Tracer diffusion in crowded narrow channels]{Tracer diffusion in crowded narrow channels. Topical review}

\author{O. B\'enichou$^{1}$, P. Illien$^{2}$, G. Oshanin$^1$, A. Sarracino$^{3,4}$ and R. Voituriez$^{1,5}$}
\address{$^1$ Sorbonne Universit\'e, CNRS, Laboratoire de Physique Th\'eorique de la Mati\`ere Condens\'ee (UMR 7600),
 4 Place Jussieu, 75252 Paris Cedex 05, France}
\address{$^2$ ESPCI Paris, PSL Research University, UMR Gulliver 7083, 10 rue Vauquelin, 75005 Paris, France}
\address{$^3$ ISC-CNR and Dipartimento di Fisica Universit\`a Sapienza Roma, Piazzale Aldo Moro 2, 00185 Rome, Italy}
\address{$^4$ Dipartimento di Ingegneria, Universit\`a della Campania ''Luigi Vanvitelli'', 81031 Aversa (CE), Italy}
\address{$^5$ Sorbonne Universit\'e, CNRS, Laboratoire Jean Perrin (UMR 8237), 4 Place Jussieu, 75005 Paris, France}
\eads{benichou@lptmc.jussieu.fr, pierre.illien@espci.fr, oshanin@lptmc.jussieu.fr, ale.sarracino@gmail.com, voituriez@lptmc.jussieu.fr}

\begin{abstract}
We summarise 
different results on the diffusion of a tracer particle in 
lattice gases of hard-core particles with stochastic dynamics, 
which are
confined to narrow channels -- single-files, comb-like structures and quasi-one-dimensional
channels with the width equal to several particle diameters. 
We show that in such geometries a
surprisingly rich,  sometimes even counter-intuitive, behaviour emerges, which is absent in unbounded systems. This is well-documented for the anomalous diffusion in single-files. Less known is the anomalous dynamics of a tracer particle in crowded branching single-files -- comb-like structures, where several kinds of anomalous regimes take place. 
In narrow channels, which are broader than single-files, one encounters a wealth of anomalous behaviours in the case where the tracer particle is subject to a regular external bias: here, one observes  
 an anomaly in the temporal evolution of the tracer particle velocity, 
 super-diffusive at transient stages, and ultimately a giant diffusive broadening of fluctuations in the position of the tracer particle, as well as spectacular multi-tracer effects of self-clogging of narrow channels. Interactions between a biased tracer particle and a confined crowded environment also produce peculiar patterns in the out-of-equilibrium distribution of the environment particles, very different from the ones appearing in unbounded systems.  
For moderately dense systems,  
a surprising effect of a negative differential mobility 
takes place, 
such that the velocity of a biased tracer particle can be a non-monotonic function of the force. In some parameter ranges, both the velocity and the diffusion coefficient of a biased tracer particle can be non-monotonic functions of the density. 
We also survey different results obtained 
for a tracer particle diffusion in unbounded systems, which will permit a reader to have an exhaustively broad picture of the tracer diffusion in crowded environments.
\end{abstract}

\vspace{2pc}

\noindent{\it Keywords}: Lattice gases, tracer diffusion, external bias, stochastic dynamics, asymmetric simple exclusion process,
negative differential mobility
\vspace{2pc}



\maketitle

\tableofcontents

\section{Introduction} 

Transport of molecules and small particles in pores, channels, or
other quasi-one-dimensional systems has attracted a great deal of
attention within several last decades \cite{0}. On the one hand,
this interest stems from the relevance of the problem to a variety of
realistic physical, biophysical and chemical systems, as well as
important applications in nanotechnology and nanomedicine, e.g., in the
creation of artificial molecular nanofilters. Few stray examples
include, e.g., transport in porins \cite{2,3,4}, through the nuclear
pores in eukaryotic cells \cite{5,6,7}, or along the microtubules
\cite{welte} and dendritic spines \cite{nim}, transport of
microswimmers in narrow pores \cite{berg,graham,mal0}, translocation
of polymers in pores \cite{sak0,sak01,sak1,sak2,sak3,tap} and their sequencing in nanopore-based devices \cite{chi}, ionic currents across
nanopores \cite{ionic1,ionic2}, ionic liquids in supercapacitors
\cite{super1,super2,super3,super4,super5,super6,super7,super8}, nano-
and microfluidics \cite{adj,muk,liot,bet} and transport of inertial particles
advected by laminar flows \cite{sar1,sar2,sar3}.  On the other hand, the
diversity and the significance of the problems emerging in this field
represent a challenging area of research for theoreticians \cite{mal}.

\begin{figure}
\begin{center}
	\includegraphics[width=9cm]{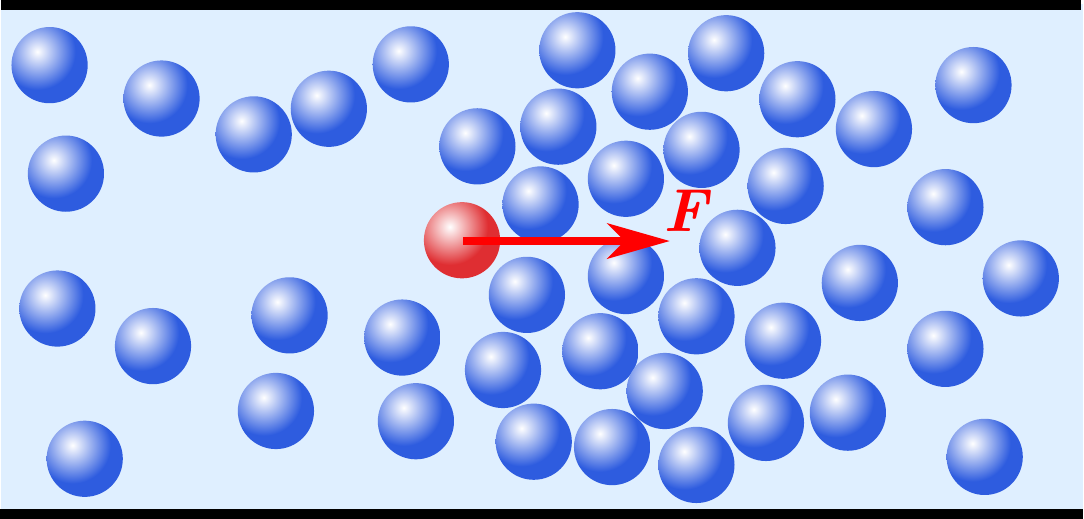}
	\end{center}
	\caption{ Tracer particle,  subject to a regular constant force $F$ pointing along the channel, in a confined bath of crowding particles.
\label{general_activemicro}
}
\end{figure}

Molecular crowding has been recognised as an important factor (see,
e.g.,
Refs. \cite{leb,der1,rzia10,kaf,bauer,chou,chou2,sasha,zilman,rzia11,ai,kafri00,mal1,kafri0,kafri,bol}
and references therein), which strongly affects the behaviour of
currents across the narrow channels, as well as the dynamics of
individual molecules or probes, which move either due to thermal
activation only, or are also subject to external constant forces.  This latter
case, i.e., dynamics of \textit{biased} particles in crowded narrow
channels is precisely the chief subject of this topical review.  More
specifically, we present here an overview of analytical and numerical
work which deals with  the behaviour observed in narrow channels -  single-files, ramified single-files (comb-like structures) 
and quasi-one-dimensional
channels with the width equal to several particle diameters -
with rigid, structureless
hard-walls, densely populated by a quiescent mixture of identical
environment particles with purely repulsive short-ranged
interactions.  
The environment particles undergo stochastic dynamics
due to thermal noise resulting, e.g., from the interactions with the
solvent present in the channel.  In addition,  there is a special
tracer particle, which has the same size and the same interaction
potential as the environment particles, but is also subject to a
regular constant force $F$ pointing along the channel (see
Fig. \ref{general_activemicro}). As one may notice, these settings
mimic the standard experimental set-up of the so-called
\textit{active} micro-rheology
\cite{muk,activ1,activ2,activ3,activ4,activ5,activ6,activ7,dullens}
and hence, the tracer particle can be thought of as the probe designed
to measure, at a molecular scale, the response of the confined passive
molecular crowding environment.

\begin{figure}
\begin{center}
	\includegraphics[width=13cm]{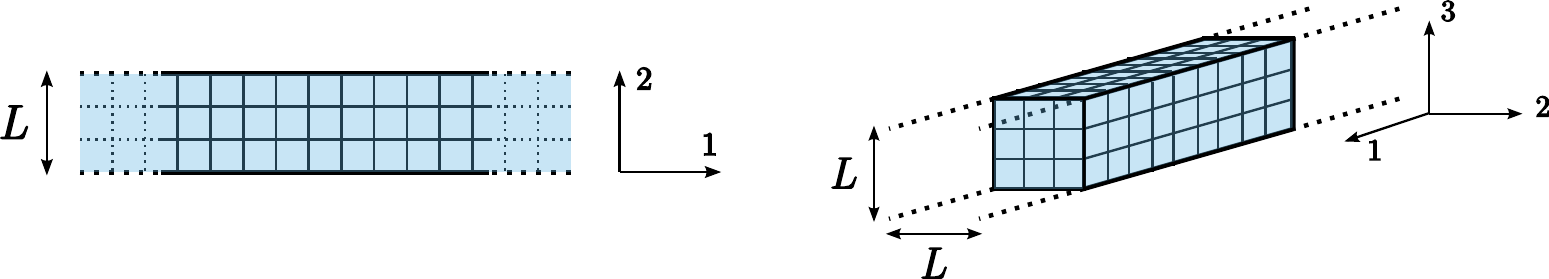}
	\end{center}
	\caption{Two-dimensional strip-like lattice with width $L = 3$ and a three-dimensional, ``capillary''-like simple cubic lattice with side $L = 3$.
\label{FIG1}
}
\end{figure}

The channel geometry on which we concentrate here is given by two
particular examples (see Fig. \ref{FIG1}) - two-dimensional ideal
strips, such as ones printed on a substrate and used in microfluidic
devices \cite{muk}, and three-dimensional ideal capillaries; both
types of systems are macroscopically large in the direction of the
applied bias, and in the perpendicular to $F$ direction
have dimensions which are \textit{comparable} to the diameters of
the environment particles. The
walls of the channels are perfect and do not contain any impurities or
geometric corrugations, and the cross-section is constant across the
channel. 
We note that a large amount of an available theoretical work concerns
lattice systems with stochastic dynamics and interactions between the
tracer particle and the environment particles, as well as between the
environment particles themselves, being a mere hard-core exclusion.
In realistic systems, of course, the particles evolve in
continuous-space and the interactions may extend beyond the
nearest-neighbouring particles, for example because  of hydrodynamic or electrostatic interactions.
This simplification, however,
permits to unveil some universal features of the dynamical behaviour
beyond the linear in the force $F$ regime. Whenever possible, we also
compare the theoretical predictions against the results of numerical
Monte Carlo or Molecular Dynamics simulations.

The purpose of this review is three-fold. First, we propose a complete
and comprehensive picture of   essentially non-equilibrium, cooperative phenomena emerging in crowded narrow
channels in response to the passage of a biased intruder. We proceed to
show that there is a large number of spectacular effects and some of them take place beyond the regime of the linear
response.   Strikingly, the behaviour appears to be distinctly different in extremely narrow channels -- the so-called single-files and in somewhat broader channels, in which the particles can bypass each other. 
Second, we note that physical situations involving a single
particle pulled by a constant force through a host medium composed of
interacting mobile particles represents the typical settings for
studying the linear-response Einstein relation (see, e.g., recent
Refs. \cite{er1,er2,er3}), which links the diffusion coefficient of this
particle in the absence of the force, and the mobility in the presence of the
latter. We comment on the validity of the Einstein relation throughout
the review
 and show that, curiously enough, it may be associated with
some subdominant contributions to the dynamics. 
Lastly, as it has been
already understood for narrow channels in which all the particles
experience a constant bias -- the so-called driven diffusive systems
(see, e.g., Refs. \cite{leb,der1,rzia10,kaf}), the explicit treatment of the multiple degrees of freedom of the environment  
is of an essential
importance. Here, as well, we proceed to show that many interesting
phenomena take place due to the interplay between these degrees of freedom and the dynamics of the tracer particle.
These effects are overlooked once one represents the environment in an
integrated, effective form and studies a Markovian Langevin dynamics of the
tracer particle in  an effective environment.

\section{Tracer diffusion in single-files, on infinite combs and on infinite (unbounded) lattices of spatial dimension $d \geq 2$}

To set up the scene, we recall here some most prominent results on the tracer particle (TP) diffusion in interacting particles systems. We do not consider as our goal to provide an exhaustive review and to pay a due tribute to all the outstanding 
contributors 
to this field - this would require too much space. Instead, we merely list below some very significant achievements relevant to the subject of our review, which were spread across many different disciplines and journals within the last five decades. This 
will permit the reader to have a somewhat broader picture.  In this way, as well, many remarkable and novel 
phenomena specific to the tracer diffusion in narrow channels will be made more apparent.

\subsection{Unbiased tracer diffusion}
\label{passive}

It was understood for a long time that diffusion of a TP  in a dynamical background formed by other interacting and mobile (randomly moving) particles is coupled in a non-trivial way to the evolution of the environment itself (see, e.g., Refs. \cite{compaan,howard,mann}). Clearly, in a sufficiently dense system, when a TP displaces in one unit of time over 
one unit of length away from its location occupied at the previous unit of time, it leaves a void of a clear space - a vacancy - behind it. 
Then, it is often more probable that the TP will return back to this location, than keeping on going away from it where its motion will be hindered by other particles. This ``anti-persistence'' of the TP motion and the circumstance whether it will be ``permitted'' by the environment particles to return back depends essentially on how fast the environment can rearrange itself and close the void. In turn, this depends on plenty of physical factors - the density of the environment, the particle-particle interactions, the temperature and the viscosity of the embedding solvent, if any. Hence, the TP diffusion coefficient is expected to acquire some dependence on all the aforementioned parameters.


\subsubsection{Single-files.} 

The single-file concept was introduced first in biophysical literature \cite{exp} to describe diffusion through pores in membranes, which are so narrow such that the initial order of particles is preserved at all times. 
In mid $60$-s it was realised that in single-files the effect of interactions with the environment can be even stronger than a mere renormalisation of the diffusion coefficient - it was shown that the single-file constraint 
 can change the very temporal evolution of the TP mean-squared displacement $\overline{X_t^2}$.  Analysing the TP dynamics in a single-file of interacting diffusions, it was discovered in Ref. \cite{harris} that $\overline{X_t^2}$ obeys, in the asymptotic limit $t \to \infty$, a slower than diffusive law, $\overline{X_t^2} \sim \sqrt{t}$, i.e., $\overline{X_t^2}$ exhibits not a linear Stokes-Einstein but an anomalous sub-diffusive growth with time. The point is that in single-files - an extreme case of narrow channels - due to the constraint that the environment particles cannot bypass each other,
  in order to explore a distance $X_t$ the TP has to involve in a cooperative motion $\sim \rho X_t$ particles of the environment, with $\rho$ being their mean density\footnote{See, e.g., the discussion after eq. (2.22) in Ref. \cite{ooshida2} for more details.}. 
 In consequence, the effective 
 frictional force exerted by the medium on the TP also scales with the distance travelled by the TP, which entails a sub-diffusive motion\footnote{The situation is, of course, different in "single-files" in contact with a vapour phase such that the environment particles can desorb from and re-adsorb back to the channel. In this case the order is not preserved and the TP undergoes a standard diffusive motion with, however, the diffusion coefficient dependent on all the system's parameters \cite{lemarchand,olla1}.}. In a way, this resembles (and in fact, it is linked on the level of the underlying mathematics) dynamics of a tagged bead in an infinitely long Rouse polymer chain; in order to explore progressively longer and longer spatial scales the tagged bead 
 has to involve in motion more and more other beads of the chain \cite{rouse}.

This remarkable result has been subsequently re-derived in different settings by using a variety of analytical techniques  \cite{levitt,rich,fedders,pincus,arratia,van,maj,bar2,bar}  (see also recent Refs. \cite{chvosta,taloni} for a more extensive review) to show that the exact 
leading asymptotic behaviour of $\overline{X_t^2}$ obeys
\begin{equation}
\label{1}
\overline{X_t^2} \underset{t\to\infty}{\sim}  \frac{\left(1 - \rho\right) a^2}{\rho} \sqrt{\frac{2}{\pi}\frac{t}{\tau}}, 
\end{equation} 
where $a$ is the lattice spacing and $\tau$ is the mean waiting time between jumps.  When $\rho \to 1$, (i.e., the space available for diffusion shrinks), the right-hand-side of eq. (\ref{1}) vanishes, since the system becomes completely blocked. In turn, in the limit $\rho \to 0$ the right-hand-side of eq. (\ref{1}) diverges, meaning that a faster growth has to take place - in this limit, of course, one expects the standard Brownian motion result $\overline{X_t^2} \sim t$ to hold. For arbitrarily small, but finite $\rho$, the result in eq. (\ref{1}) will describe the ultimate long-time behaviour, while the diffusive law will appear as a transient.

The time dependence of  eq. (\ref{1})
was also observed experimentally (see, e.g., Refs. \cite{hahn,wei,meer,lin} and also Refs. \cite{chvosta,taloni}), which is not surprising given its universal nature.  
The TP dynamics in single-files still 
represents a challenging playground for testing different analytical techniques and probing other properties: convergence of the distribution of $X_t$ to a Gaussian \cite{arratia}, explicit form of the distribution of $X_t$ at arbitrary, not necessarily  large times in dense systems \cite{active},  the TP dynamics in the presence of a slower-than-diffusive environment \cite{ralf},
large deviations properties \cite{dhar,krap0,der,krap,ima} and emerging correlations \cite{ooshida2,m1,m2,ooshida,m4,m5}. 
Surprisingly, some unexpected
 features of this seemingly exhaustively well-studied process keep on being revealed:  it was shown recently \cite{barkai} (see also Refs. \cite{krap0,kundu2,barkai3}) that the preservation of the initial order implies, in fact, an \textit{infinite} memory of the process 
 on the specific initial conditions -- in systems, in which the particles of the environment are initially out-of-equilibrium (also called a"quenched" initial condition),
the growth of the mean-squared displacement of the TP proceed a factor of $\sqrt{2}$ slower, than in systems with an equilibrium  (also called "annealed" or uniform) initial distribution of the environment particles\footnote{Note that this result has been obtained earlier in Ref. \cite{barkai2}, but was not explicitly mentioned in this work. }.  In Ref. \cite{krap0} such a  dichotomy in dynamics under annealed and quenched initial conditions 
has been also shown to persist for higher-order cumulants of the TP position, for which it becomes even more striking. 

\begin{figure}[t]
\begin{center}
	\includegraphics[width=0.6\textwidth]{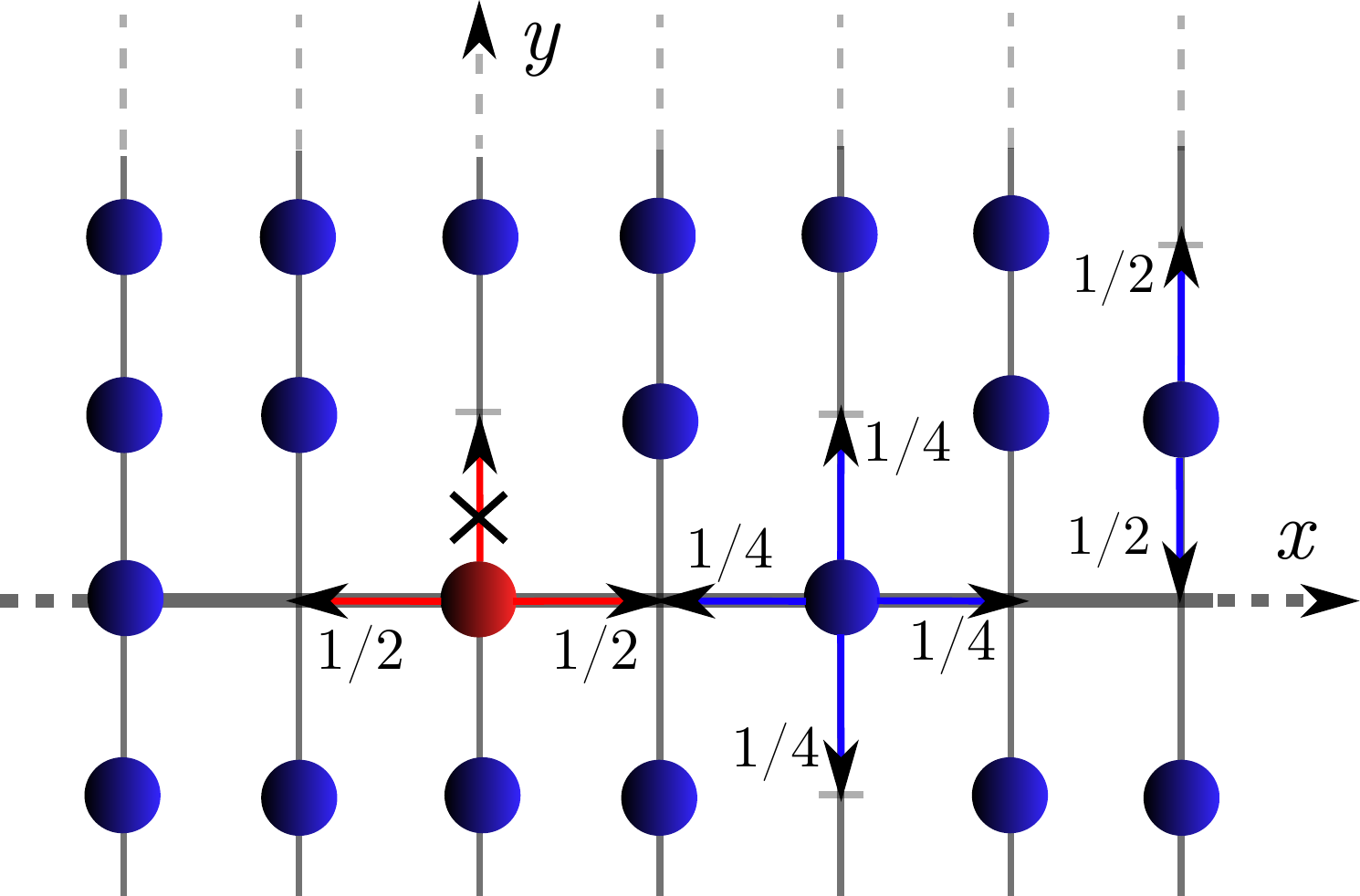}
	\caption{Unbiased tracer particle (red filled circle) on a comb-like lattice populated by hard-core environment particles (blue filled circles). The $x$ axis is a backbone, whereas the orthogonal lines are the infinitely long teeth of the comb.
\label{FIG2}
}
\end{center}
\end{figure}

\subsubsection{Comb-like structures.}

 Before we pass from single-files 
to the behaviour on unbounded regular lattices of dimension $d$ higher than $1$, it seems relevant to focus on some geometrically ``intermediate'' situation, in which a variety of anomalous diffusions emerge. 
Recently Ref. \cite{comb} studied the dynamics of a TP in the presence of hard-core environment particles on a comb-like lattice, i.e., a one-dimensional backbone (see Fig. \ref{FIG2}) connected at each site to a tooth  -- an infinite one-dimensional lattice, with passages of particles between the teeth being allowed only along the backbone. Dynamics of a single isolated particle in such  a system has been widely studied in the past as a toy model of dynamics in a geometrically disordered system \cite{comb1}. It is well known that the mean-squared displacement of a single TP (in the absence of the environment particles) obeys, at sufficiently long times, $\overline{X_t^2} \sim \sqrt{t}$ \cite{comb1,weisshavlin}, i.e., the TP moves sub-diffusively along the $x$-axis due to progressively longer and longer 
excursions on the teeth. 
Ref. \cite{comb} thus analysed a natural extension of this model by adding stochastically moving environment particles and considering two situations: (a) the environment particles can move everywhere, i.e., both along the backbone and along 
the teeth, while the TP can go along the backbone only, and (b) both the TP and the environment particles are permitted to go everywhere.

At the first glance, the case (a) seems to be  very close to the situation discussed in Refs. \cite{lemarchand,olla1}, i.e., a one-dimensional lattice attached to a reservoir of particles. Hence, one may expect essentially the same ``diffusive'' behaviour, i.e.,  $\overline{X_t^2} \sim D t$, where the diffusion coefficient $D$ captures the combined effect of the geometry and dynamics. In reality, the behaviour appears to be more complicated. Ref. \cite{comb} focused on densely populated systems and analysed the 
TP mean-squared displacement in the leading in the density $\rho_0 = 1 - \rho$ of vacancies order, in the limit $\rho_0 \to 0$. It was shown that, when the vacancies are initially uniformly spread  across the comb with mean density $\rho_0$, the mean-squared displacement of the TP along the $x$-axis obeys, for times $t$ sufficiently large but less than a certain large cross-over time $t_1 \sim 1/\rho_0^4 \ln^8(1/\rho_0)$:
\begin{equation}
\label{comba}
\lim_{\rho_0 \to 0}\frac{\overline{X_t^2}}{\rho_0} = \frac{1}{2^{5/4} \Gamma(7/4)} t^{3/4} \,,
\end{equation} 
where $\Gamma(x)$ is the gamma-function, which crosses over for $t > t_1$ to the diffusive behaviour of the form
\begin{equation}
\label{combb}
\lim_{t \to \infty}\frac{\overline{X_t^2}}{t} = \rho_0^2 \left(\ln\frac{1}{\rho_0}\right)^2 \,.
\end{equation} 
The results in eqs. (\ref{comba}) and (\ref{combb}) reveal two interesting features: First, it is the existence of  
a transient regime with a sub-diffusive behaviour, which is more extended in time the denser the system is.    Second, in the ultimate diffusive regime, the diffusion coefficient of the TP exhibits a rather unusual non-analytic dependence on the density of vacancies.

Reference \cite{comb} described an interesting out-of-equilibrium situation when the vacancies are all initially placed on the backbone only, with a linear density $\rho_0^{(lin)} \ll 1$.  Interestingly enough,  in this case
the mean-squared displacement of the TP grows sub-linearly
\begin{equation}
\label{combc}
\overline{X_t^2} = \frac{\rho_0^{(lin)}}{2^{7/4} \Gamma(5/4)} t^{1/4} \,,
\end{equation}
for any (sufficiently large) time $t$, without any crossover to an ultimate diffusive regime. We note, as well, that  
all \textit{even} cumulants of the TP displacement 
follow the dependences in eqs. (\ref{comba}) and (\ref{combb}) in case of a uniform distribution of vacancies
on the comb, and the dependence  in eq. (\ref{combc}) in case of their placement on the backbone only, which signals that the distribution of the TP position is a Skellam distribution (see, Ref. \cite{comb} for more details). This means that the distribution converges to a Gaussian at long times, with an appropriately rescaled TP position.

Lastly, in the case (b) when both the environment particles and the TP can go along the teeth, the following dynamical scenario has been predicted and confirmed through an extensive numerical analysis \cite{comb}. For an extended time interval $0 \ll t \ll t^{(1)} \sim 1/\rho_0^2$, one has $\overline{X_t^2} \sim t^{3/4}$, in which regime the TP has not had enough time to explore any given tooth  because its excursions were hindered by the environment particles. This regime is followed, on the time interval $t^{(1)} \ll t \ll t^{(2)}$ (precise form of $t^{(2)}$ was not provided in Ref. \cite{comb}), by a rather unusual dependence $\overline{X_t^2} \sim t^{9/16}$, associated with the large-$t$ tail of the corresponding distribution of the time spent by the TP on a given tooth. Ultimately, for $t \gg t^{(2)}$, one finds again $\overline{X_t^2} \sim t^{3/4}$. Note that the coincidence of the dynamical exponent $3/4$ characterising the initial and the final stages is occasional; the underlying physics is completely different. Note, as well, that at all stages the growth of the TP mean-squared displacement proceeds faster than in case of a single isolated TP on an empty comb ($\overline{X_t^2} \sim t^{1/2}$).   This is a consequence of interactions with the environment particles, which do not permit the TP to enter too often into the teeth and also to travel too far within each tooth.

\subsubsection{Unbounded lattices with $d \geq 2$.}

 On lattices of spatial dimension $d \geq 2$,  the ultimate long-time dynamics of the TP 
is diffusive\footnote{See, however, Ref. \cite{galanti,gosh,greb} for the non-diffusive 
behaviour in case of an inhomogeneous crowding.}, such that $\overline{X_t^2} = 2 d D t$. As we have already remarked, here the main issue is the calculation of the effective diffusion coefficient which embodies 
the full dependence on both the rates of the TP and of the environment particles,  their density and the type of the lattice (if any), on which the evolution takes place. 
It was fairly well understood that calculation of the diffusion coefficient is a genuine many-body problem, which is unsolvable in the general case and one has to resort to either density expansions or some other approximations, verified by numerics.
Numerous approaches have been developed and a large number of different results, both analytical and numerical, have been obtained (see, e.g., Refs. \cite{fedders2,nakazato1,koiwa,kehr1,nakazato2,koiwa2,tahir1,van2,tahir2,henk1,henk2} for a few stray examples,  Ref. \cite{binder} for some early 
review and Ref. \cite{pigeon} for  a more recent summary).   
We focus below particularly on two approaches, proposed in Refs. \cite{nakazato1} and \cite{henk1,henk2}, respectively, which will be used in what follows for the analysis of the TP dynamics in narrow channels. 

In Ref. \cite{nakazato1},
the self-diffusion constant of a tracer on regular lattices partially populated by 
identical hard-core particles has been analysed using an approximate approach, based on a perturbative expansion in powers of $\rho (1-\rho)$, which is 
exact at the two extrema of the volume fraction, $\rho=0$ and $\rho=1$. This approach gives, in particular, for an unbounded $d$-dimensional hyper-cubic lattice the following expression for the TP diffusion coefficient
\begin{equation}
\label{2}
D = \frac{(1 - \rho) a^2 }{2 d \tau} f(\rho, \omega) \,,
\end{equation}
where $1/\tau$ is the jump frequency of the TP, $a$ is the lattice spacing, and $\omega = \tau^*/\tau$ with $1/\tau^*$ being the characteristic jump frequency of the environment particles. Lastly,   $f(\rho, \omega)$ is the correlation factor which
is explicitly given by
\begin{equation}
f(\rho, \omega) = \frac{\left(\omega (1 - \rho) + 1 \right) \left(1 - \alpha\right)}{\omega (1 - \rho) + 1 - \alpha \left(1 + \omega (1 - 3 \rho)\right)} \,,
\end{equation}
with $\alpha$ being one of the Watson integrals \cite{zucker}
\begin{equation}
\label{alpha}
\alpha = \frac{1}{(2 \pi)^d} \int^{\pi}_{- \pi} dk_1 \ldots \int^{\pi}_{-\pi} dk_d \frac{\sin^2\left(k_1\right)}{\sum_{j=1}^d \left(1 - \cos\left(k_j\right)\right)} \,.
\end{equation}
This latter factor
relates $D$ to the propagator of a special type of random walk on a $d$-dimensional hyper-cubic lattice. For $d=1$, one has  $\alpha \equiv 1$ and hence, $D$ vanishes, as it should, since here we enter into the realm of a single-file diffusion. 
On comparing the analytical predictions against available at that time results of Monte Carlo simulations in $d=3$ systems, the authors demonstrated that their approximation
 gives a good interpolation formula in between of the two extrema.  In Refs. \cite{ol1,ol2} and \cite{ol3}, the expression in eq. (\ref{2}) has been generalised, using a similar approach based on a decoupling of the TP-particle-particle correlation function into a product of two TP-particle correlation functions, for situations when the density $\rho$ of the environment particles 
  is not explicitly conserved (unlike, e.g., the case of a monolayer in a slit-pore, confined between two solid surfaces) but the particles undergo continuous exchanges with a reservoir maintained at a constant chemical potential.  Refs. \cite{ol1,ol2} focused specifically on such a situation in two dimensions, i.e., a monolayer on top of a solid surface exposed to a vapour phase, while Ref. \cite{ol3} presented an explicit expression for the TP diffusion coefficient in three-dimensions, which case represents some generalised model of a dynamic percolation.  The resulting expressions for the diffusion coefficient are quite cumbersome and we do not present them here, addressing an interested reader to the original papers \cite{ol1,ol2,ol3}. We note parenthetically that in case when exchanges with the vapour are permitted, dynamics becomes diffusive and $D$ is well-defined \cite{lemarchand,olla1} even in single-files, since the initial order is destroyed.
  
 A different line of thought has been put forth in Refs. \cite{henk1,henk2}, which studied the TP dynamics in a very densely 
 populated system of hard-core particles, assuming that a continuous  
 ``reshuffling'' of the environment occurs due to the dynamics of
the ``vacancies'' (i.e., vacant sites, present on the lattice at mean density $\rho_0=1 - \rho$), rather than of the particles themselves. Focussing on  such a ``vacancy-assisted'' dynamical model on a two-dimensional square lattice of a unit lattice spacing, the authors studied first 
 the case of just a single vacancy, which moves in the discrete time $n$, $n=0,1,2, \ldots$, 
by exchanging its position, at each tick of the clock, with one of the neighbouring particles (including the TP, once it appears on the neighbouring site), chosen at random. In such a two-dimensional model, any particle including the TP performs
 concerted random excursions and can, in principle, travel to infinity from its initial position. By counting all possible paths which bring the TP to position ${\bf X}$ at time moment $n$, the authors were able to calculate exactly the full probability distribution $P_n({\bf X})$ of finding the TP at this very site at time moment $n$. Surprisingly, this distribution appears to be non-Gaussian even in the limit $n \to \infty$ 
and the mean-squared displacement $\overline{{\bf X}_n^2}$ shows a striking sub-diffusive behaviour in the leading in $n$ order \cite{henk1} 
\begin{equation}
\label{3}
\overline{{\bf X}_n^2} = \frac{\ln(n)}{\pi (\pi - 1)} \,,
\end{equation}
with a very non-trivial numerical amplitude (which depends, of course, on the lattice structure). 

Reference \cite{henk2}, focused on the case of a very small density of vacancies. 
Discarding the events when any two vacancies appear at the adjacent  sites (whose probability is of order of $\rho_0^2$), 
an analogous expression for  $P_n({\bf X})$ has been derived, which does converge to a Gaussian in the limit $n \to \infty$. The TP dynamics in this case appears to be diffusive, since now the TP displacement is a sum of many independent events. In consequence,
 the mean-squared displacement, at sufficiently large $n$, obeys
\begin{equation}
\label{4}
\overline{{\bf X}_n^2} = \frac{\rho_0}{\pi - 1} n \,,
\end{equation}
 with, again, a very non-trivial numerical amplitude, 
 which depends on the lattice structure. 
 We emphasise that this result is only valid in the linear in the density of vacancies $\rho_0$ order.  There exist, 
 of course, $\rho_0$-dependent corrections to the diffusion coefficient. Expanding the expression in eq. (\ref{2}) in Taylor series in powers of $\rho_0$ and retaining only the leading term in this expansion, one may straightforward verify that it coincides with the expression in eq. (\ref{4}). 
 
The result in eq. (\ref{4}) has been subsequently generalised in Ref. \cite{puzzle} for the situations where the TP interacts with the environment particles in a different way, than the particles interact between themselves, which situation mimics, e.g., dynamics of an intruder (e.g., an Indium atom) in the close-packed upper most layer of atoms in a metal (e.g., ${\rm Cu}$), in the presence of one or a few naturally existing defects of packing - the vacancies \cite{gastel1,gastel2}. The expression derived in Ref. \cite{puzzle} revealed an interesting non-monotonic dependence of the self-diffusion coefficient on the strengths of these interactions.  Vacancy-assisted dynamics on different types of lattices has been amply discussed in Refs. \cite{sh1,sh2,sh3,sh4,sh5,sh6}.
 
 By assuming the validity of the Einstein relation, a generalisation of the expression in eq. (\ref{4}) for a hyper-cubic lattice of an arbitrary dimension $d$ can be obtained from the expression for the TP mobility (see the Supplementary Information to Ref. \cite{geom}). 
 This gives the following expression
 \begin{equation}
 \label{5}
 D = \frac{\rho_0}{2 d} \frac{1 - \alpha}{1 + \alpha} \,,
 \end{equation} 
 where $\alpha$ is defined in eq. (\ref{alpha}), $\tau^*= \tau=1$ and $a=1$.
 We note that eq. (\ref{5}) coincides with the leading term in the expansion in Taylor series in powers of $\rho_0$ of the expression in eq. (\ref{2}).

\subsection{Biased tracer diffusion}

We turn to the situations when the particles of the environment still experience no other regular force except for random thermal ones, but the TP is subject to some regular constant 
force $F$ pointing along the $X$-axis. It should be emphasised that the presence of a constant bias exerted on the TP only does not merely add another dimension to the parameter space, but crucially changes the dynamical behaviour in the system under study. Within the context of the aforementioned experimental technique - micro-rheology \cite{muk,activ1,activ2,activ3,activ4,activ5,activ6,activ7}   - the situations described in Sec. \ref{passive} correspond to the so-called \textit{passive} micro-rheology, when the TP, as well as the environment particles move solely due to thermal forces and the system is in thermal equilibrium. On the contrary, when the TP is biased by an external force, the situation corresponds to the typical settings of an \textit{active} micro-rheology. Here, 
the TP entrains in motion the environment particles appearing in its vicinity, which produces 
micro-structural changes in the host medium and brings it out of equilibrium. 
This means that the system becomes characterised by highly asymmetric stationary density profiles of the environment particles, in the frame of reference moving with the TP. Except for the single-files, which still exhibit a very particular dynamics (i.e., the TP creeps, instead of moving ballistically, $\overline{X_t} \sim \sqrt{t}$, and the density profiles around it do not converge to a stationary form), in higher dimensions 
the TP attains a constant velocity $V$ which results from an interplay between the jamming of the medium, produced by the TP, and diffusive re-arrangements of the environment. Various aspects of this problem have been
 studied experimentally (see, e.g., Refs. \cite{furst1,furst2,bech1}), 
 using a combination of numerical simulations and theoretical approaches    (see, e.g., Refs. \cite{mode1,zia}) and by extensive numerical simulations (see, e.g., Refs. \cite{reich1,win,horbach}). An inverse problem in which the TP is kept fixed (e.g., by an optical tweezer), while the environment particles are subject to a constant force has also been rather extensively studied \cite{dzu,furst3,kol1,kol2}.
Below we present some most important 
results for single-files and unbounded, infinite in all direction systems, obtained for the models of lattice gases with stochastic dynamics.

\subsubsection{Single-files.}
\label{single}

A model with a biased TP travelling in a  lattice gas of particles with symmetric hopping probabilities has been studied in Ref. \cite{driven1} in the particular limit when the force is infinitely large, such that the TP performs a totally directed random walk in the direction of the applied bias, 
constrained by hard-core interactions with the environment particles.  It was shown in Ref. \cite{driven1} that  when the TP jumps instantaneously to the neighbouring lattice site on its right, as soon as it becomes vacant, and does not jump in the opposite direction, its mean displacement obeys at sufficiently long times
\begin{equation}
\label{mean}
\overline{X_t} = \gamma  \, \sqrt{2 D_0 t} \,,
\end{equation}
where $D_0$ is the Stokes-Einstein diffusion coefficient of the tracer particle in the absence of the environment particles and
the amplitude $\gamma$ is defined implicitly as the solution of the following transcendental equation
\begin{equation}
\gamma I_{+}(\gamma) = \rho_0 \,, \,\,\, I_+(\gamma) = \sqrt{\frac{\pi}{2}} \exp\left(\frac{\gamma^2}{2}\right) \left(1 - {\rm erf}\left(\frac{\gamma}{\sqrt{2}}\right)\right) \,,
\end{equation}
with ${\rm erf}(x)$ being the error function. Remarkably, the TP does not move ballistically, i.e., does not  have a constant velocity, but rather creeps. This happens because the displacement of the TP becomes effectively hindered by the environment particles; at each step to the right, the TP "pumps" a vacancy to the left, such that the density of the environment particles in the phase in front of the TP effectively grows. 
In turn, 
there appears a sort of a traffic jam in front of the TP, which also grows in size in proportion to the distance $X_t$ travelled by the TP. This means that the system is out  of equilibrium (in contrast to the situations described in Sec. \ref{passive}) at any time, the density profile of the environment is a step-like function propagating to the right away of the TP such that more and more of the environment particles are entrained by the TP in a directed motion. Viewing $X_t$ as a random process described by a Langevin equation, one may argue \cite{driven1} that in this situation there is
an effective frictional force exerted on the TP due to the appearance of the propagating traffic jam in front of it, which grows in proportion to $X_t$.
 
In Refs. \cite{driven2,driven3} a more complicated situation has been analysed when the force $F$ exerted on the TP is finite, such that it may perform jumps in both directions\footnote{See also Ref. \cite{driven0} for some applications in biophysics,  Refs. \cite{driven01,driven02,driven00} for a relevant analysis within the harmonisation approximation and Ref. \cite{huv2} for the solution in a similar mathematical set-up.}. It was shown in Refs. \cite{driven2,driven3} that the TP mean displacement still has a form in eq. (\ref{mean}), but now the amplitude $\gamma$ is determined by a more complicated transcendental equation
\begin{equation}
\label{A}
\fl \,\,\,\,\,\,\,\,\,\,\,\,\,\,\,\,\,\,\,\,\, \,\,\,\,\,\,\,\,\,\,\, \,\,\,\,\,\,\,\,\,\,\, \,\,\,\,\,  \left(\gamma I_+(\gamma) + \frac{e^{-\beta F} - \rho_0}{1 - e^{-\beta F}}\right)\left(\gamma I_-(\gamma) + \frac{1 - \rho_0 e^{-\beta F} }{1 - e^{-\beta F}}\right) = \frac{\rho_0^2 e^{-\beta F}}{\left(1 - e^{-\beta F}\right)^2} \,,
\end{equation}
where $\beta$ is the reciprocal temperature and
\begin{equation}
I_-(\gamma) = \sqrt{\frac{\pi}{2}} \exp\left(\frac{\gamma^2}{2}\right) \left(1 + {\rm erf}\left(\frac{\gamma}{\sqrt{2}}\right)\right) \,.
\end{equation}
Refs. \cite{driven2,driven3} have also 
demonstrated that the density profiles of environment particles in front of and past the TP attain a stationary form in variable $x = (X - \overline{X_t})/\overline{X_t}$ characterised by a dense region in front of the TP and a depleted region in its wake. In consequence, in the laboratory frame, the environment particles progressively accumulate in front of the TP,  and the size of the depleted region past the TP, which is devoid of particles, also grows progressively in time. Overall, the biased TP entrains in a directed motion the entire system, which is quiescent in the absence of the external bias. This is quite evident for the particles in front of the TP, but less trivial for the particles in its wake. The explanation is, however, rather simple and relies on one of the results presented in Ref. \cite{arratia}; Namely, it was shown that in case of an asymmetric layout of the environment particles around an unbiased TP, such that they are initially 
present past the TP and are absent in front of it, the mean displacement of the TP follows $\overline{X_t} = \sqrt{t \ln(t)}$, i.e., an unbiased TP moves faster than $\sqrt{t}$ due to an additional logarithmic factor, which stems from the pressure exerted on the TP by the particle phase behind it. This means that in case of a symmetric placement of the environment particles around the biased 
TP, once the TP goes away from the phase behind it, the closest to the TP particle 
sees the void space in front of it and starts to move as $\sqrt{t \ln(t)}$ catching eventually the TP, which moves as $\sqrt{t}$.

The results in eqs. (\ref{mean}) and (\ref{A}) permit to make the following, conceptually important conclusion:  Define a generalised, time-dependent mobility $\mu_t$ of the TP in the presence of the force as
\begin{equation}
\label{mobility}
\mu_t = \lim_{F \to 0} \frac{\overline{X_t}}{F t}  \,,
\end{equation}
and the time-dependent diffusivity $D_t$ in the absence of the external force as $D_t = \overline{X^2_t}/2 t$. Our aim is now to check if the Einstein relation $\mu_t = \beta D_t$ holds. Note that, first, for standardly defined mobility and the diffusion coefficient, which requires going to the limit $t \to \infty$, the Einstein relation trivially holds since $0 = 0$. A legitimate and non-trivial question is if it holds as well for time-dependent $\mu_t$ and $D_t$.  
Calculating $D_t$ from eq. (\ref{1}) and $\mu_t$ - from eqs. (\ref{mean}) and (\ref{A}), one realises that the Einstein relation holds exactly in the case when the initial distribution of the environment particles is random \cite{driven2,driven3}, and does not hold  when the latter is regular \cite{barkai}, which is a  non-trivial result. 
The validity of the Einstein relation in single-file systems has been also investigated via Molecular Dynamics simulations in models of granular (inelastic) particles in Refs. \cite{villa1,villa2}.

Within the last decade many other characteristic features of biased TP diffusion in single-file systems have been studied. In particular, 
the mean displacement of a biased TP has been analysed in case when the initial distribution of the environment particles has a shock-like profile:  
a higher density of the environment particles from the left from the TP than from the right of the TP, and the TP is subject to a constant force $F$ pointing towards the higher density phase, (which mimics an effective attraction towards this phase) 
 \cite{driven4}. It was shown that $\overline{X_t}$ obeys eq. (\ref{mean}) in which the amplitude
$\gamma$ is determined implicitly as the solution of a quite complicated transcendental equation (see, Ref. \cite{driven4}). Interestingly enough, the TP not always moves in the direction of the force but only  when the latter exceeds some threshold value $F_c$. 
For $F < F_c$, the TP moves against the force and the high density phase expands. When the force is exactly equal to this threshold value, $\gamma = 0$ such that the mean displacement vanishes; in this particular case, the mean-squared displacement is not zero, $\overline{X_t^2}  \sim \sqrt{t}$ with a prefactor dependent on the densities in both phases (see, Ref. \cite{driven4}). 

In recent Ref. \cite{driven5} dynamics of a pair of biased TPs in a single-file has been analysed in case when they experience action of the forces pointing in the opposite directions, such that the forces tend to separate the TPs. It was shown that in this situation, however, the TPs do indeed travel in the opposite directions only when the forces exceed some critical value; otherwise, the pair of TPs remains bounded.  

The full distribution of the TP position and as well as its cumulants have been calculated exactly for an arbitrary time 
 in the leading order in the density of vacancies $\rho_0$ \cite{active}, upon an appropriate generalisation of the approach developed in Refs. \cite{henk2}. The TP position distribution has been also determined in case of an arbitrary particles density within a certain decoupling approximation \cite{dist1,dist2}, for the situation when exchanges of particles with a reservoir are permitted. It was shown that in the asymptotic limit $t \to \infty$ the distribution converges to a Gaussian. Interestingly enough, it was realised that in the presence of a bias the diffusion coefficient can be a non-monotonic function of the environment particles density $\rho$.  Lastly, the emerging correlations in such systems have been analysed in Refs. \cite{kundu3,kundu4}. 
 
We close this subsection by noting that the models with a biased TP evolving in a single-file of particles with symmetric hopping probabilities appear also in different contexts (seemingly unrelated to the TP diffusion). 
In particular, they describe the dynamics of the 
front of a
propagating precursor film which emanates from a spreading liquid droplet \cite{spr1,spr2,spr3,spr4,spr5} or a time evolution of a
 boundary of a monolayer on a solid surface \cite{spr6}.

\subsubsection{Density profiles of the environment particles.}

On unbounded lattices of spatial dimension $d \geq 2$, the biased TP 
attains ultimately a constant velocity $V$ along the direction of the force, whose dependence  
on system's parameters is rather non-trivial and will be discussed below. At the same time, the TP 
alters the environment, 
involving 
in a directional motion 
some of the environment particles and hence, 
drives the environment out of equilibrium - it is no longer homogeneous and some density profiles emerge.  
In the frame of reference moving with the TP, these density profiles attain a steady-state form meaning that, (in contrast to the case of single-files),
here the entrainment is only partial,  in the
sense that upon an encounter with the TP an environment particle 
 travels alongside the TP
for some random time, leaving it afterwards and being replaced
by another particle. The total amount of the entrained environment particles stays
constant, on average, in time. 

The emerging steady-state 
density profiles have been studied in detail in Refs. \cite{ol1,ol2,ol3} for square and simple cubic lattices. It was realised that, in general, in front of the TP and in the direction perpendicular to the force, the approach of the perturbed (local) 
density to its average value is exponential, with the characteristic length dependent on the coordinates, such that they are asymmetric. 
In the wake of the TP,   
the form 
of these profiles 
depends essentially on whether one deals with the situation in which the number of the environment particles is explicitly conserved, or the system is exposed to some vapour phase and the environment particle may desorb from and re-adsorb into the system - in this latter case the particles number fluctuates in time around some average value. Then, the local density past the TP approaches the unperturbed value exponentially fast, but the characteristic length protrudes over much larger scales than in front of the TP. The situation is very different in the conserved number case. Here, the local density approaches its unperturbed value as a power-law,  meaning that the environment remembers the passage of the TP over very large temporal and spatial scales. The approach of the local density $\rho(\lambda)$ to the mean density $\rho$ obeys \cite{ol1,ol2,ol3}
\begin{equation}
\label{dens}
\rho - \rho(\lambda) \sim \frac{C}{\lambda^{(d+1)/2}} \,,
\end{equation}  
where $\lambda$ is the distance past the TP. The amplitude $C$ in this decay law has been also calculated in Refs. \cite{ol1,ol2,ol3}.

We close this subsection with the following two remarks:

First, we note that such a spectacular  
behaviour of the density 
of the environment particles in the wake of a biased TP has been also seen 
 in various continuous space
settings (see, e.g., Refs. \cite{furst1,zia,reich1,horbach,demery}), 
 in granular media \cite{dau} and also in colloidal experiments (see, e.g., Refs. \cite{furst2,furst3}).
 
Second, we note that an interesting phenomenon may take place in situations where there is not a single biased TP, but there are a few or even a concentration of them. The point is that an out-of-equilibrium environment may mediate long-ranged mutual interactions between intruders, which are non-reciprocal and violate Newton's third law (see, e.g., Refs. \cite{low,soft6}).  When several TPs are present, each of them will perturb the distribution of the environment particles in its vicinity and in order to minimise the micro-structural changes in the environment and to reduce the dissipation, they will start to move collectively. Stochastic pairing of two biased 
TPs has been observed in simulations in a model of a two-dimensional lattice gas \cite{soft1},  in experiments in colloidal suspensions \cite{furst4,soft5},
and also theoretically predicted for the relative motion of two TPs in a nearly-critical fluid mixture, due to emerging critical Casimir forces \cite{cazimir}.   Formation of string-like clusters of TPs, or ``trains'' of TPs, 
which reveals an emerging effective attraction between them has been evidenced 
for situations with a small concentration of TPs \cite{reich3,ladadwa,soft2}. Lastly, in case when the concentration of the TPs is comparable to that of the environment particles, the TPs arrange themselves in lanes \cite{rex1,rex2,glanz,soft3,soft4}. This self-organisation of the TPs resembles spontaneous lane formation in binary mixtures of oppositely charged colloids (see, e.g., Ref. \cite{wysocki}), in dusty complex plasmas  \cite{rel2}  and also for such seemingly unrelated systems as pedestrian counter flows \cite{rel3}.

\subsubsection{Mean velocity of the tracer particle.}

The functional dependence of the TP ultimate velocity $V$ on physical parameters has been studied in details 
within the last two decades.  In the most general case, these parameters are
the magnitude of the external force, the density of the environment, the rates of the particles exchanges with a vapour phase, and the jump rates of both the TP and of the environment particles. 
Several analytical 
approaches have been used in this analysis - 
an analytical technique 
based  on a decoupling of correlations between the tracer particle and two given environment particles (see, Refs. \cite{ol1,ol2,ol3}), 
and also a different  approach based 
on an appropriate generalisation of the vacancy-assisted dynamical model of Refs. \cite{henk1,henk2},  in 
which  
 a bias exerted on the TP has been taken into account explicitly \cite{beni1}.

References \cite{ol1,ol2,ol3} studied the functional dependence of $V$ on all physical parameters, including the exchange rates of the environment particles with the vapour phase (which define the value of the density $\rho$), in the general case of unequal jump rates of the TP and of the environment particles. The resulting expressions appear very cumbersome, and we address the reader to the original papers. Here we merely mention some features of the expressions obtained in Refs. \cite{ol1,ol2,ol3}:

First, in the limit of very small forces, $V$ attains a physically meaningful form
\begin{equation}
\label{Vel}
V = \xi^{-1} F \,,
\end{equation} 
which can be thought of as an analogue of the Stokes formula with $\xi$ being the friction coefficient. This 
signifies that the frictional force exerted on the TP by the environment is \textit{viscous} at small values of $F$. In turn, the friction coefficient can be decomposed into two contributions \cite{ol1,ol2,ol3}: a mean-field one, proportional to an effective density of voids, and the second one which has a more complicated form and stems from the cooperative behaviour emerging in such a system - de-homogenisation of the environment by the TP and formation of some asymmetric density profiles of the environment particles around the TP, which we will discuss later on. 

Second, it was observed in Refs. \cite{ol1,ol2,ol3} that in the systems with an explicitly conserved particle density,  the mobility of the TP exactly equals
$\beta D$ with $D$ defined in eq. (\ref{2}) meaning that the Einstein relation holds. 

Third, we note that the expressions obtained in Refs. \cite{ol1,ol2,ol3} have been re-examined recently in Ref. \cite{neg} and it was realised that for the environments which evolve on time scales larger than the one which defines the TP jump rates, the terminal velocity $V$ is a \textit{non-monotonic}\footnote{See also Sec. \ref{VF} below for the analogous behaviour of the velocity versus force in strip-like geometries.} function of the force $F$. The velocity first increases with an increase of $F$, as dictated by the linear response theory, reaches a maximal value and then, upon a further gradual increase of $F$, starts to decrease. In consequence, the differential mobility of the TP, defined beyond the linear response regime, becomes negative. In Ref. \cite{neg}, a full phase chart
has been presented indicating the region in the parameter space in which such a non-monotonic behaviour emerges.
 We note that the phenomenon of the so-called negative differential mobility  is a direct consequence of an out of equilibrium dynamics. Various facets of this intriguing phenomenon  have been studied in the past and it has been also observed in many realistic physical systems \cite{mo,eich,jack,sel,turci,urna1,urna2,bai1,sar1,sar2,civ1,negative1,urna}.  
 
A different approach to calculation of the terminal velocity of a biased TP in very densely crowded environments has been pursued in Ref. \cite{beni1}, which generalised the analytical technique originally developed in Refs. \cite{henk1,henk2}, over the case when the TP is subject to an external constant force $F$. Reference \cite{beni1} first focused on the case of a single vacancy-mediated dynamics evolving in discrete time $n$ and calculated exactly the mean displacement  of a biased TP on an infinite square lattice to get
\begin{equation}
\label{displacement}
\overline{X}_n = \frac{1}{\pi} \frac{\sinh\left(\beta F/2\right)}{\left(2 \pi - 3\right) \cosh\left(\beta F/2\right) + 1} \ln(n) \,,
\end{equation}
i.e., the TP displaces anomalously slowly (logarithmically) with time. Accordingly, 
the time-dependent mobility $\mu_n$ (see eq. (\ref{mobility})) of the biased TP obeys
\begin{equation}
\label{mobb}
\mu_n = \frac{\beta}{4 \pi (\pi - 1)} \frac{\ln(n)}{n} \,.
\end{equation}
Determining from eq. (\ref{3}) the effective diffusivity of the TP in the absence of the force, $D_n = \overline{X_n^2}/4 n$, one arrives at a surprising conclusion \cite{beni1}: the Einstein relation $\mu_n = \beta D_n$ holds exactly even for such an anomalously confined diffusion! The full distribution function of the TP position 
has been also determined in Ref. \cite{beni1}  and it was shown that it \textit{does not} converge to a Gaussian as $n \to \infty$, similarly to the case without an external bias \cite{henk1}.

Reference \cite{beni1} extended this analysis over the case when vacancies are present on a square lattice 
at very small density $\rho_0$, to find that in the lowest order in the density of vacancies the mean displacement of 
the biased TP obeys
\begin{equation}
\label{nnn}
\overline{X}_n = \frac{\sinh\left(\beta F/2\right)}{\left(2 \pi - 3\right) \cosh\left(\beta F/2\right)+1} \rho_0 n \,,
\end{equation}
such that the TP now moves ballistically with a finite velocity $V = \overline{X}_n/n$. Inspecting the mobility of the TP and the diffusion coefficient of the TP in eq. (\ref{4}), one realises that the Einstein relation holds \cite{beni1}. The distribution of $X_n$ has also been determined in Ref. \cite{beni1} and it was shown that the latter converges to a Gaussian as $n \to \infty$.

In more recent Ref. \cite{geom}, the result in eq. (\ref{nnn}) has been generalised, using essentially the same technique as in Ref. \cite{beni1}, to calculate the mean velocity of a biased TP on a $d$-dimensional hypercubic lattice (including the simple cubic one, $d=3$) in the presence of a small density of vacancies. The result of Ref. \cite{geom} (see the Supplementary Materials to this paper) reads
\begin{equation}
\label{a0}
V = a_0 \rho_0 \,, \,\,\, a_0 = \frac{(1 - \alpha) \sinh\left(\beta F/2\right)}{\left(d (1 + \alpha)-1+\alpha\right) \cosh\left(\beta F/2\right)+1 - \alpha}  \,,
\end{equation}
where the parameter $\alpha$ is defined in eq. (\ref{alpha}). For $d =2$, eq. (\ref{a0}), multiplied by $n$, reduces to the expression in eq. (\ref{nnn}). For $d = 1$, the parameter $\alpha$ becomes equal to $1$, such that $V$ vanishes, as it should. 

\subsubsection{Variance of the tracer particle displacement.} 

We concentrate on the time-evolution and the density dependence of the variance $\sigma^2_x$ of a biased
TP displacement in the direction of the force $F$, $\sigma^2 = \overline{X_t^2} - \overline{X}_t^2$, 
on a two-dimensional square and a three-dimensional simple cubic lattices.  
This question has been raised only recently and several rather surprising results have been obtained. Note that in the absence of the force $F$, the variance simply becomes $\sigma^2_x = 2 d D t$, where the forms of the diffusion coefficient have been discussed above.

We start with the case of a square-lattice, densely populated by the environment particles, and a biased TP, and focus on the vacancy-assisted dynamics in discrete time $n$, which was put forth in Refs. \cite{henk1,henk2}. 
First observation of a ``strange'' behaviour of the variance $\sigma_x^2$ has been made in Ref. \cite{variance1}, in which it was predicted analytically and verified numerically that $\sigma_x^2 \sim n \ln(n)$. 
Surprisingly, it appeared that in a dense system the spread of fluctuations in the TP position is (weakly) super-diffusive. This question has been further inspected in Ref. \cite{variance2} and the physical mechanism responsible for such a super-diffusive growth of fluctuations has been explained - it was shown that it emerges due to correlations between the successive jumps of the TP, which originate from interactions with a single vacancy returning to the TP position many times and carrying it in the direction of the force.

Deeper analysis of this model 
has been presented in Ref. \cite{geom}, 
and it was shown that this super-diffusive regime is only a transient one which prevails up to times of order of $1/\rho_0^2$. Not only the leading term but also the correction terms to the leading behaviour have been calculated, to provide the following exact expression
\begin{equation}
\label{var}
\lim_{\rho_0 \to 0} \frac{\sigma_x^2}{\rho_0} \sim a_0 n \Bigg[\frac{2 a_0}{\pi} \left(\ln(n) + \ln 8 + C -1 +  \frac{\pi^2 (5 - 2 \pi)}{(2 \pi - 4)}\right) + \coth\left(\frac{\beta F}{2}\right) \Bigg] \,,
\end{equation}
where $C$ is the Euler-Mascheroni constant, the parameter $a_0$ is defined in eq. (\ref{a0}) and the omitted terms 
do not depend on $n$, i.e., are unimportant for sufficiently large $n$. Recalling that $a_0 \sim F$ for small forces, eq. (\ref{var}) implies that the coefficient in front of the leading super-diffusive term is proportional to $F^2$, meaning that super-diffusion emerges beyond the linear response. Note that using eq. (\ref{a0}) at small $F$ and eq. (\ref{var}) at $F=0$, the Einstein relation is shown to be satisfied.

For times greater than $1/\rho_0^2$, the weakly super-diffusive regime crosses over to a standard diffusive growth of the variance, which emerges due to renewal events - arrivals of other vacancies to the location of the TP -  which fade out  the memory about interactions with a single vacancy and make successive jumps of the TP uncorrelated.
In this ultimate regime, one finds that in the leading in $\rho_0$ order the variance obeys \cite{geom}
\begin{equation}
\lim_{n \to \infty} \frac{\sigma_x^2}{n} \sim a_0 \rho_0 \Bigg[\coth\left(\frac{\beta F}{2}\right)
+2 a_0 \left(\frac{2}{\pi} \ln\left(\frac{1}{a_0 \rho_0}\right) + \frac{1}{\pi} \ln 8 + \frac{\pi (5 - 2 \pi)}{2 \pi - 4} \right) \Bigg] \,.
\end{equation}
Here, again, one observes that the effective diffusion coefficient acquires some additional terms which are quadratic with the force in the limit of small forces. Note that for moderate forces, however, these terms can become large due to the factor $\ln(1/\rho_0)$. The validity of the Einstein relation in this ultimate regime is again ensured by the term $a_0 n \coth(\beta F/2)$.

For the TP dynamics on a simple cubic lattice, densely populated by the environment particles, the growth of the variance of the TP displacement obeys at all time scales \cite{geom}
\begin{equation}
\lim_{\rho_0 \to 0} \frac{\sigma_x^2}{\rho_0} \sim a_0 n \left[\coth\left(\frac{\beta F}{2}\right) + A a_0\right] \,,
\end{equation}
where $A$ is a numerical constant. 
Hence, the spread of fluctuations in three- (and actually, higher-) dimensions is diffusive but the effective diffusion coefficient contains, as well, quadratic with the force terms in the limit of small applied forces. The validity of the Einstein relation is again ensured by the term $a_0 n \coth(\beta F/2)$.

Lastly, we discuss the observations made in Ref. \cite{variance3}, which analysed the density-dependence of the variance of the TP displacement in a system of the environment particles evolving on a discrete lattice of an arbitrary dimension 
in continuous time $t$. The approach of Ref. \cite{variance3} was based on a decoupling of the TP-particle-particle correlation function, similar to the one employed in Refs. \cite{ol1,ol2,ol3}, and the results were verified numerically and in the high- and low-density of particles limits, in which exact solutions are available \cite{geom,franosch1,franosch2,franosch3}.  In particular, it was shown in Ref. \cite{variance3} that in the ultimate regime ($t \to \infty$) the effective diffusion coefficient can be a non-monotonic function of the density, for forces which exceed a certain threshold value, well above the linear response regime, and is a monotonic function of $\rho$ for $F$ below the threshold. Moreover, in the case when the jump rate of the environment particles is substantially lower than the one characterising the jumps of the TP, this effective diffusion coefficient can be also a non-monotonic function of the force $F$ - a behaviour we have already observed for the velocity of a biased TP.

\section{Biased tracer diffusion in narrow channels}

We turn eventually to the dynamics of a biased TP in narrow channels - strip-like or capillary-like lattices depicted in Fig. \ref{FIG1}.  These lattices have width $L$, (with $L$ being an integer), in case of a strip, and a cross-section $L \times L$ in case of a capillary, and are of an infinite extent in the direction of the force $F$ acting on the TP only. As above, we will focus on the density profiles of the environment particles around a steadily moving TP,  its velocity  $V$, 
and the variance $\sigma^2_x$ of the TP displacement at time $t$. We proceed to show that in such confined geometries the time-evolution of the system exhibits many remarkable features, which are absent on infinite lattices with $d \geq 2$.

\subsection{Density profiles of the environment particles}

The density profiles of the environment particles around a steadily moving TP in strip-like and capillary-like geometries have been studied analytically, using a decoupling of the TP-particle-particle correlation functions into a product of two TP-particle correlation functions, 
and also numerically in recent Refs. \cite{channels1,channels2}. It was realised that, similarly to the case of lattices which have an infinite extent in all directions, these density profiles attain a stationary form in the frame of reference moving with the TP. These profiles are asymmetric and
are characterised by a dense,
traffic jam-like region in front of the TP and a depleted by the environment particles region in the wake of the TP. The traffic jam-like region is more pronounced than in an infinite space \cite{channels2}, which is not counter-intuitive. 
This means that the frictional force exerted by the host medium on the TP is bigger and hence, the velocity is smaller in strips and capillaries than on infinite square and simple cubic lattices. The density, as a function of the distance from the TP in the direction perpendicular to the force, is described by an exponential function with the characteristic length that  depends in a rather complicated way on the width  $L$ of the strip or the cross section of a capillary. When $L \to \infty$, the results in Refs. \cite{ol1,ol2,ol3} are recovered.  The form of the stationary density profiles past the TP is more complicated: there is some characteristic length scale $\lambda^*$, $\lambda^* \sim V L^2/D$, where $V$ is the TP terminal velocity and $D$ - the diffusion coefficient of the environment particles, such that at distances $\lambda < \lambda^*$ the evolution of the density profile in the wake of the TP is described by a power-law, eq. 
(\ref{dens}), where $d=2$ for strips and $d=3$ for the capillaries. Beyond this distance, the approach of the density to its unperturbed value $\rho$ becomes exponential, with the characteristic length dependent on $L$ and diverging when $L \to \infty$. This overall behaviour has been confirmed numerically for lattice systems using Monte Carlo simulations \cite{channels1,channels2} and also for off-lattice, strip-like geometries with hard-core particles with Langevin dynamics \cite{channels2}. Returning to the discussion of the entrainment properties, here, as in infinitely large systems, we encounter a \textit{partial} entrainment of  the environment particles by the biased TP but the amount of the entrained particles is bigger.

To close this subsection, we briefly mention recent observations made in Ref. \cite{kuster}, in which dynamics of several biased TPs in presence of the environment particles in strip-like geometries has been analysed via extensive Molecular Dynamics simulations. It was observed that for sufficiently large driving forces, a strong clogging of the channel takes place - 
the biased TPs cluster together such that the particles of the traffic jam region in front of them 
cannot circumvent the cluster. In consequence, the cluster of the TPs propagates in proportion to $\sqrt{t}$, i.e., the velocity of each individual TP vanishes, such that one is led back to an effectively single-file behaviour (see Sec. \ref{single}).

\subsection{Mean velocity of a biased tracer particle}

In this subsection we present available analytical and numerical results on the mean velocity $V$ of a biased TP in strip-like and capillary-like geometries. We note that the time-evolution of the TP displacement shows in such a confinement a bit richer behaviour than on infinite lattices with spatial dimension $d \geq 2$. A salient feature here is an emergence of a plateau-like behaviour at intermediate time scales.

\subsubsection{Transient mean velocity. }

Before we proceed to the discussion of the forms of $V$ in narrow channels, we briefly recall one of the results of Ref. \cite{henk2}, which concerns the form of the diffusion coefficient $D_L$ for self-diffusion of an unbiased TP on an infinitely long 
strip of width $L$, densely populated by the environment particles. In Ref. \cite{henk2}, within the model with the  vacancy-assisted dynamics, the following exact, in the first order in the density $\rho_0$ of vacancies, result has been presented
\begin{equation}
\label{L}
D_L = \frac{\rho_0}{4} \frac{1 - \alpha'}{1+ \alpha'} \,,
\end{equation} 
where 
\begin{equation}
\label{alpha'}
\alpha' = \frac{1}{4} \lim_{z \to 1^-} \left[P(0|0;z) - P(2 {\bf e}_x|0;z)\right] \,, 
\end{equation}
with ${\bf e}_x$ being a unit vector in the positive $x$-direction and 
$P({\bf r}|0;z)$ - the generating function associated with the propagator of a symmetric random walk on a two-dimensional strip of width $L$, starting at the 
origin and ending at position ${\bf r} = (n_1,n_2)$,
\begin{equation}
\label{z}
P({\bf r}|0;z) = \frac{1}{2 \pi L} \sum_{k=0}^{L-1} \int^{\pi}_{-\pi} dq \frac{e^{- i n_1 q - 2 i \pi k n_2/L}}{1 - z \left(\cos q + \cos(2 \pi k/L)\right)/2} \,.
\end{equation}
Note that $\alpha'$ here has essentially the same meaning as $\alpha$ in eq. (\ref{alpha}), except that now it is defined for confined geometries.

This situation has been re-examined in Ref. \cite{geom} in case of a TP  biased by a constant external force, to show that for both two-dimensional strips ($d=2$) and three-dimensional capillaries ($d=3$) the velocity of the TP is given, in the same order in the density of vacancies,
by 
\begin{equation}
\label{V}
\tilde{V} = \rho_0 a_0' \,, \,\,\, a_0' = \frac{(1 - \alpha') \sinh\left(\beta F/2\right)}{\left(1 + (2 d - 1) \alpha' \right) \cosh\left(\beta F/2\right) + 1 - \alpha'} \,,
\end{equation}
where $\alpha'$ is defined in eq. (\ref{alpha'}) (with the factor $1/4$ replaced by $1/(2 d)$) and with $P({\bf r}|0;z)$ given in eq. (\ref{z}) in case of two-dimensional strips, and by a bit more complicated expression describing the propagator of a random walk in case of three-dimensional capillaries (see Ref. \cite{geom} for more details).  One may directly verify, upon comparing the diffusion coefficient defined in eq. (\ref{L}) and the mobility deduced from eq. (\ref{V}) with $d=2$, that the Einstein relation holds.

\subsubsection{Terminal mean velocity. }

A surprising observation has been made subsequently in Ref. \cite{anomaly}, in which it was realised that $\tilde{V}$ in eq. (\ref{V}) does not describe the true velocity of the biased TP in the limit $n \to \infty$, but only a constant velocity appearing at some transient stage, for times less than the cross over time $t_x \sim 1/\rho_0^2$ (which is however quite large when $\rho_0 \ll 1$). The point is that in the underlying derivation the limit $\rho_0$ is taken first and then the analysis focuses on the asymptotic behaviour in the limit $n \to \infty$. It appears, however, that due to some very subtle circumstances these limits do not commute and, in essence,  
the velocity $\tilde{V}$ in eq. (\ref{V}) corresponds to the temporal regime when the TP interacts with just a single vacancy.
Recall that we have already encountered  this phenomenon while describing the behaviour of the variance of the TP displacement on an infinite square lattice.  
The key difference consists in the fact that for a \textit{fixed} small $\rho_0$ the random walk performed by any of the vacancies between two successive visits to the TP position is a \textit{biased} random walk, in the frame of reference moving with the TP, due to the interactions of the TP with other vacancies. In consequence, to get the expression for the ultimate, terminal velocity $V$, one has to examine the general expressions taking the limit $n \to \infty$ first, and only then to concentrate on the leading behaviour in the limit $\rho_0 \to 0$. In doing so, it was shown in Ref. \cite{anomaly} that for $n \gg 1/\rho_0^2$ the true terminal velocity 
obeys
\begin{equation}
\label{alpha''}
V = \rho_0 a_0^{''}\,, \,\,\, \frac{1}{a_0^{''}} = \frac{1}{a'_0} + \frac{2 d}{1 - \alpha'} \frac{1}{L^{d - 1}} \,,
\end{equation}
where $a_0'$ is defined in eq. (\ref{V}) and $\alpha'$ - in eq. (\ref{alpha'}). 

The exact expression in eq. \ref{alpha''}
 permits to draw several important conclusions, which all were also confirmed in Ref. \cite{anomaly} by extensive Monte Carlo simulations:\\
 \begin{itemize}
 \item in strips and capillaries with a finite $L$ the true terminal velocity is lower than the velocity $\tilde{V}$ at the transient stage;\\
 \item this effect does not exist on infinite lattices when $L \to \infty$, even for $d=2$ when the variance shows two distinct behaviours; \\
 \item the effect is more pronounced in strip-like geometry than in the capillaries;\\
 \item it requires quite long times $(\sim 1/\rho_0^2)$ to observe the terminal velocity;\\
 \item the jump of the velocities $\tilde{V} - V \sim F^2$ for small forces $F$,  meaning that this anomaly takes place beyond the linear response regime;\\Ö
 \item the latter also ensures that the Einstein relation holds on both transient and terminal stages.\\
 \end{itemize}

Lastly, the analysis in Ref. \cite{anomaly} culminated at the derivation of a complete expression for the time evolution of the TP mean displacement, which is valid at an arbitrary discrete time $n$:
\begin{equation}
\frac{\overline{X}_n}{\rho_0 n} \sim g\left(\left(a_0^{''}\right)^2 \rho_0^2 n\right) \,,
\end{equation} 
with 
\begin{eqnarray}
g(\tau) &=& a'_0 \Bigg[\frac{b}{b^2-1} \left({\rm erf}\left(\sqrt{\tau}\right) + \frac{e^{-\tau}}{\sqrt{\pi \tau}}\right) + \frac{b}{2} \frac{b^2+1}{\left(b^2-1\right)^2} \frac{{\rm erf}\left(\sqrt{\tau}\right)}{\tau} - \nonumber\\
&-& \frac{1}{b^2 - 1} + \frac{1}{\tau} \left(\frac{b}{b^2-1}\right)^2 \left(e^{(b^2-1) \tau} {\rm erf}\left(b \sqrt{\tau}\right) -1\right) \Bigg] \,,
\end{eqnarray}
where $b = a'_0/a_0^{''} - 1$. This scaling function reproduces correctly both regimes and has been numerically verified in Ref. \cite{anomaly} for both strip-like and capillary-like geometries. 

\subsubsection{General force-velocity relation. }
\label{VF}

\begin{figure}[t]
\begin{center}
	\includegraphics[width=0.4\textwidth]{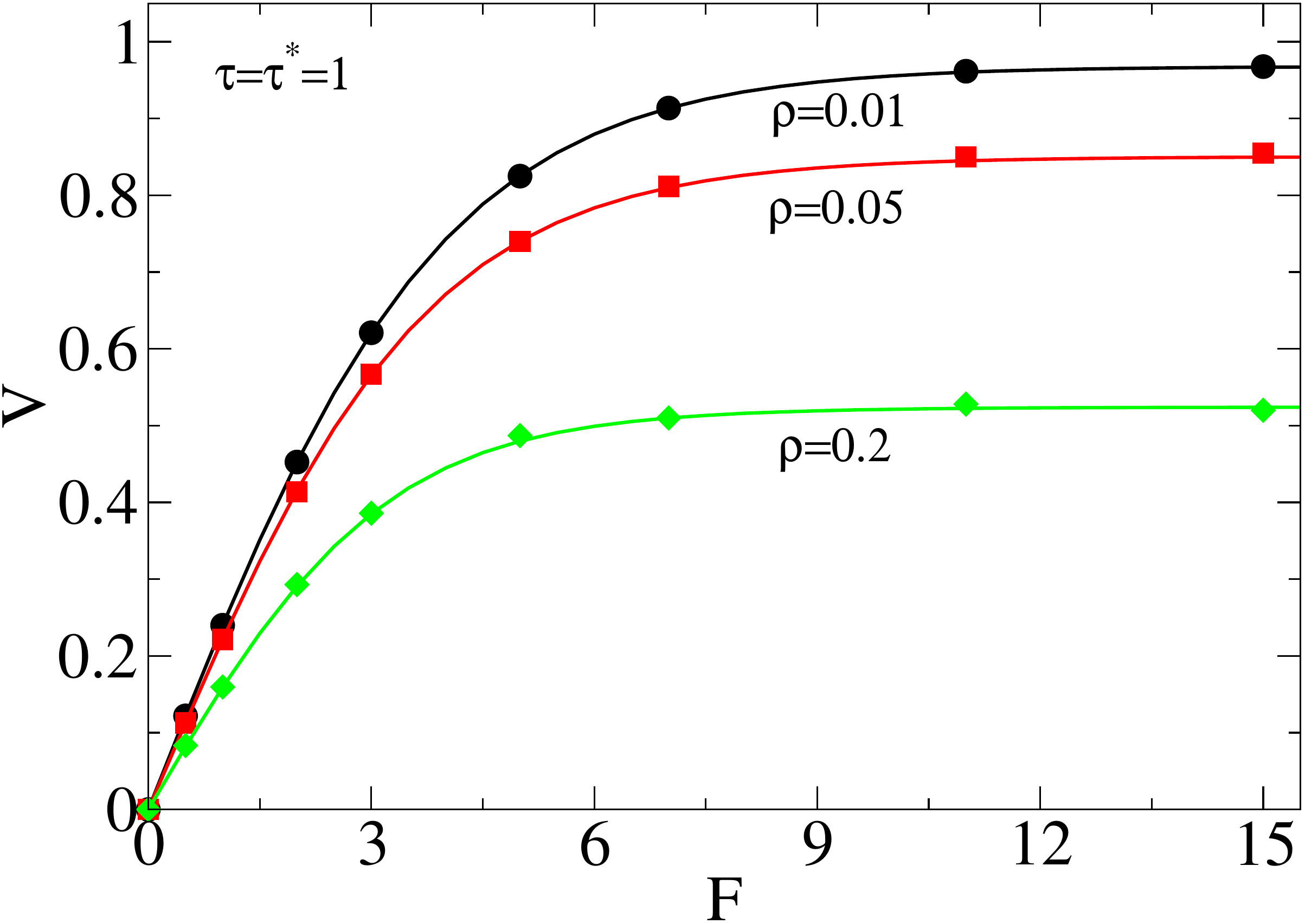}
	\includegraphics[width=0.4\textwidth]{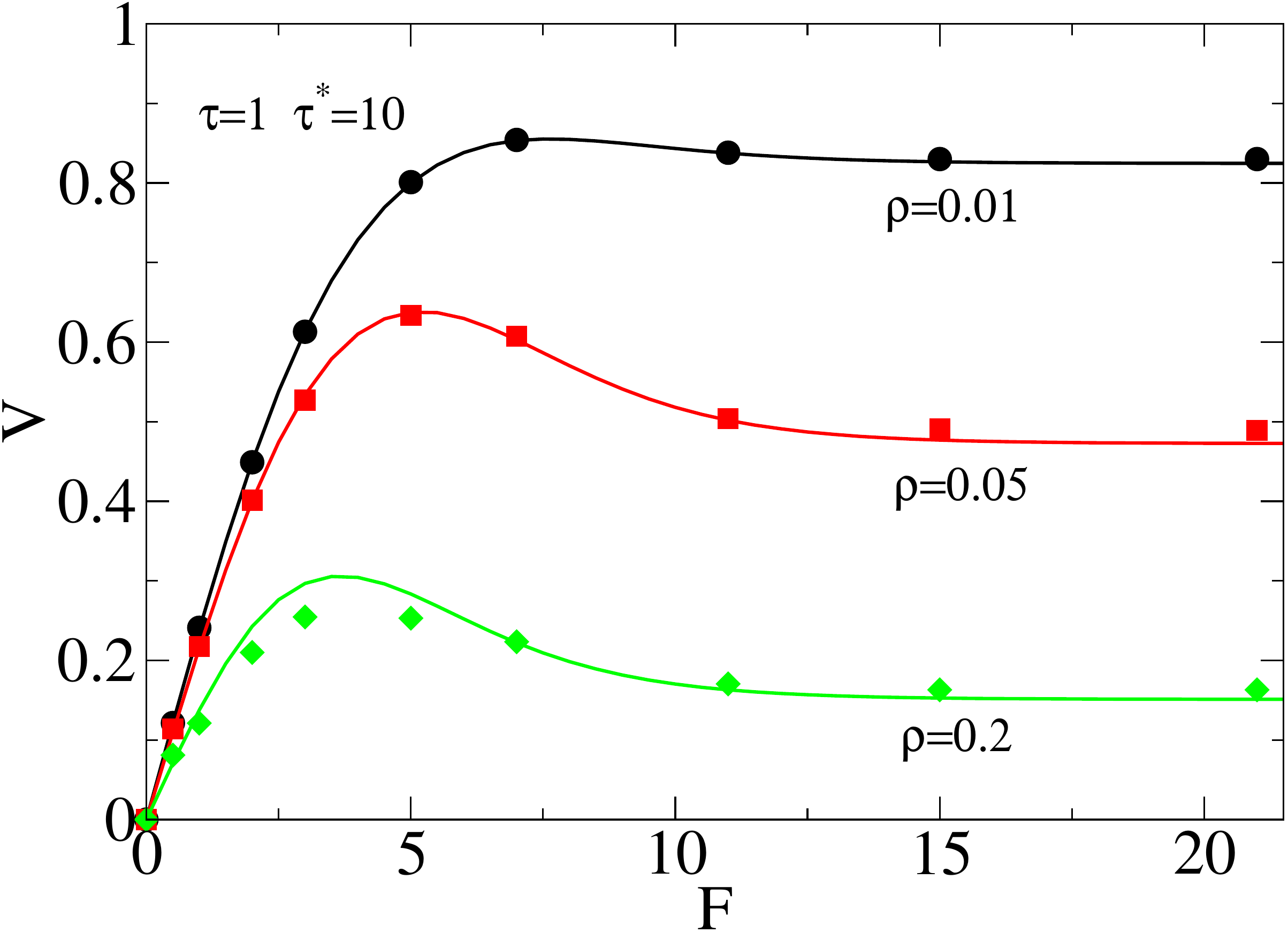}
	\caption{Strip-like geometry with $L=3$. Analytical predictions (continuous curves) in Ref. \cite{channels1} against the results of numerical simulations (symbols). Left panel: the terminal velocity $V$ versus force $F$ for different densities and $\tau= \tau^*=1$. Right panel:  Analogous results for $\tau=1$ and $\tau^*=10$ (slowly evolving environment).
\label{FIG3}
}
\end{center}
\end{figure}

General force-velocity relation for a biased TP moving in crowded strip-like or capillary-like geometries, 
with an arbitrary density $\rho$ of the environment particles 
and different characteristic jump times of the TP, $\tau$, and of the environment particles, $\tau^*$, 
has been derived in Ref. \cite{channels1} using the decoupling approximation. In Ref. \cite{channels1}, the terminal velocity has been obtained as an implicit solution of a rather complicated non-linear equation involving matrix determinants, which we do not present here. We merely depict in Fig. \ref{FIG3} the curves $V$ versus $F$ for
a particular case of strip-like geometries with the width $L=3$, for 
 several densities $\rho$ of the  environment particles, fixed characteristic jump-time of the TP, $\tau =1$, and two different jump-times of the environment particles, $\tau^* = 1$ and $\tau^* = 10$. We observe that for the former case, when the characteristic jump-times of the TP and of the environment particles are equal to each other, the terminal velocity $V$ is a monotonic function of the force $F$, which was also noticed in Ref. \cite{channels2}. Conversely,  in case of a slowly varying environment, e.g., for $\tau^* = 10$, the velocity $V$ exhibits a strongly non-monotonic behaviour. For small $F$, a linear dependence $V \sim F$ is observed, as dictated by the linear response, then, $V$ reaches a peak value $V^*$ and then gradually decreases upon a further increase of $F$. This means that the negative differential mobility phenomenon takes place also in confinement.

\subsection{Variance of the biased tracer particle displacement: from super-diffusion to giant diffusion}

The first, to the best of our knowledge, analysis, which focused 
specifically on the time-evolution of the variance  of the displacement of a biased TP 
in a crowded environment in a capillary-like 
geometry was performed in Ref. \cite{win}. In standard settings of active micro-rheology,  the authors tracked in
Molecular Dynamics simulations trajectories of a single TP pulled by a constant, sufficiently large force $F$
in a glass-forming Yukawa mixture. 
To a great surprise,  it was realised that the variance of the displacement of this driven TP along the direction of the force 
grows in proportion to $t^z$ 
with $z$ very close to $1.5$,
i.e.,  the fluctuations in the position exhibit a super-diffusive broadening. In Ref. \cite{win}, it was suggested that the physical mechanism underlying this remarkable broadening of fluctuations in the
nonlinear scaling regime is associated with a hopping motion of the TP from cage to cage: it was indeed observed 
that the TP in such a system is localised for
some time in a cage formed by the surrounding environment particles
before it quickly moves to the next cage. Whereas in the
cage the motion of the TP exhibits only small anisotropies, it becomes strongly anisotropic with respect to the
motion out of the cages. It was also argued that the 
essential feature for the motion
of the TP in the direction of force is a broad waiting
time distribution, which was shown to exhibit a fat tail. 
The authors therefore suggested a scenario which is reminiscent of a certain
class of trap models,  relating it  to the glass-forming properties of the binary Yukawa mixture and the ensuing complex energy landscape.

This analysis has been revisited in the numerical simulations performed in Ref. \cite{geom}, which studied dynamics of a TP  pulled by a constant force in mono-disperse dense fluids 
in a continuum strip-like geometry. More specifically, in Ref. \cite{geom}, Molecular Dynamics simulations of 
colloidal fluids and Brownian Dynamics 
of granular fluids (with only one value of the restitution parameter) have been performed, evidencing a super-diffusive broadening of fluctuations, $\sigma_x^2 \sim t^{1.5}$, over a very extended transient regime, which ultimately crossed over at longer times to a diffusive growth of fluctuations in the TP displacement, $\sigma_x^2 \sim t$, with a prefactor - an effective diffusion coefficient - being much bigger than the diffusion coefficient of the environment particles.  The mono-disperse colloidal fluids, as well mono-disperse granular fluids are not glass-formers, and there are no complex energy landscape in these systems, but still the super-diffusive regime exists. This naturally questions the scenario proposed in Ref. \cite{win} and calls for a different explanation.

In Ref. \cite{geom} a different physical mechanism 
has been proposed, which entails - for dense lattice gases in narrow channels (both strip-like and capillary-like) - a long-lived super-diffusive broadening of fluctuations with the dynamical exponent $3/2$ 
which is ultimately followed by a crossover to the diffusive regime with a giant diffusion coefficient. This mechanisms is based on the diffusive motion of rare (in dense systems or at a sufficiently low temperatures) vacancies - defects of packing - which are present at very low density\footnote{Note that there is no direct relation between $\rho_0$, defined for lattice systems, and the specific free volume in continuous-space settings. } $\rho_0$ and permit the particles to move by exchanging their positions with the vacancies. 
In other words, all the environment particles and the TP are completely blocked most of time and make a single jump only when any of the  vacancies arrives to their location.
It was claimed in Ref. \cite{geom} that the transient super-diffusive regime
exists for times less than $1/\rho_0^2$ and is 
associated with the interactions of the TP with just a single vacancy, which may be the initially closest one. This vacancy, once it arrives to the TP position,  is certain to return to it again many times, and the vacancy indeed keeps
on returning to the TP  carrying the latter along the direction of the force. It was emphasised in Ref. \cite{geom} 
 that the statistics of these multiple returns of a given vacancy to the instantaneous position of the TP
 in quasi-one-dimensional systems 
 is  compatible with the observations made in Ref. \cite{win}. In consequence, the displacements 
 of the TP are correlated due to multiple interactions with this single vacancy which entails the super-diffusive broadening of fluctuations. 
 At longer times, however, other vacancies will appear at the location of the TP and carry it along the direction of the force. These ``renewal'' events de-correlate 
  the consecutive  displacements of the TP, such that the fluctuations in its position will ultimately grow diffusively.
  
An exact solution of the model with vacancy-assisted dynamics 
on strip-like or capillary-like lattices has been provided in Ref. \cite{geom}. It was  shown that 
 in the linear in the density of vacancies order, the variance $\sigma^2_x$ of the TP position along the direction of the force $F$    obeys for sufficiently large (but less than $1/\rho_0^2$) discrete time $n$ :
\begin{equation}
\label{cap1}
\sigma_x^2 \sim \frac{8 (a'_0)^2 \rho_0}{3 \sqrt{\pi} L} n^{3/2}  + O\left(n\right)
\end{equation}
for strip-like geometries and for the capillary-like ones one has
\begin{equation}
\label{cap2}
\sigma_x^2 \sim \frac{4 \sqrt{2} (a'_0)^2 \rho_0}{\sqrt{3 \pi} L^2} n^{3/2} + O\left(n\right)  \,,
\end{equation}
where the omitted terms grow linearly with $n$,  
parameter $a'_0$ is defined in eq. (\ref{V}) and is associated with the TP's transient mean
 velocity in these systems.

These results demonstrate the super-diffusive growth of fluctuations with an exponent $3/2$, which is in a perfect agreement 
with the observations made in numerical simulations and also show that this striking behaviour emerges beyond the linear response regime. Indeed, the prefactor in front of $n^{3/2}$ is proportional to the second power of $F$ for small values of $F$. Interestingly, the validity of the Einstein relation is ensured here by the subdominant in time terms in eqs. (\ref{cap1}) and (\ref{cap2}), as verified in Ref. \cite{geom}.

In turn, for times $n \gg 1/\rho_0^2$, the variance $\sigma_x^2$ attains a different form, which was also calculated exactly in Ref. \cite{geom} (see also recent Ref. \cite{variance3}):
\begin{equation}
\label{last}
\sigma_x^2  \sim \frac{2}{L^{d-1}} \left[\frac{1}{a'_0} + \frac{2 d}{(1-\alpha') L^{d-1}}\right]^{-1} \, n \,,
\end{equation}   
where $\alpha'$ is defined in eq. (\ref{alpha'}), $d=2$ in case of the strip-like geometries and $d=3$ for the capillary-like ones. 

Remarkably, the effective diffusion coefficient in eq. (\ref{last}) is independent of the density of vacancies $\rho_0$. The physical origin of such a behaviour has been explained in Ref. \cite{geom} and is a very peculiar feature of transport in crowded narrow (quasi-one-dimensional) channels. Note, as well, that the diffusion coefficient of the environment particles is linearly proportional to $\rho_0$
in the low vacancy density limit. This means that the diffusion coefficient of the biased TP can be, in fact, orders of magnitude bigger than the diffusion coefficient of the environment particles. In consequence, one may claim that the super-diffusion in eqs. (\ref{cap1}) and (\ref{cap2}) paves the way to giant diffusion. We also note a recent Ref. \cite{huv}, which studied dynamics of a biased passive tracer in a diffusive environment and revealed a similar super-diffusive transient behaviour.

\section{Summary and outlook}

In this review we have summarised numerous achievements made within the last several decades in the field of unbiased and biased tracer diffusion in lattice gases with stochastic dynamics on infinite in all directions lattices, and in crowded,  infinitely long narrow channels with an embedded lattice structure  and periodic boundary conditions in the directions perpendicular to the bias. For dense environments, a very good understanding of different facets of the dynamical behaviour has been achieved via a combined effort of exact theoretical and extensive numerical analysis, and in some instances, i.e., for the single-file systems, also via an experimental analysis. 
It was realised that in narrow channels  surprisingly rich,  sometimes quite unexpected and even counter-intuitive behaviours emerge, which are absent in unbounded systems. This is, of course, a well-known anomalous diffusion in single-files, both in the absence and in the presence of an external bias acting on the tracer particle only, in which the initial order in placement of the particles is preserved at all times. Strikingly, in narrow channels which are broader than single-files permitting thus for an effective mixing of the particles, as well, a wealth of anomalous behaviours takes place : these are the velocity anomaly of the biased tracer particle, super-diffusive at transient stages and ultimate giant diffusive broadening of fluctuations of the position of the tracer particle, spectacular multi-tracer effects resulting in self-clogging,  and a variety of dynamic laws appearing in ramified channels, as exemplified here by the dynamics of a tracer particle on crowded comb-like structures. Interactions between biased tracer particles and a confined crowded environment also produce peculiar non-equilibrium patterns in the distribution of the environment particles, very different from the ones appearing in unbounded systems. 
For moderate density systems, similar effects have been evidenced via theories based on an appropriate decoupling of third-order correlation functions and numerical simulations. In particular, a surprising effect of a negative differential mobility has been unveiled showing that the ultimate velocity of a biased tracer particle can be a non-monotonic function of the force. In some parameter ranges, both the velocity and the diffusion coefficient of a biased tracer particle can be non-monotonic functions of the density. Such a behaviour pertains, as well, in unbounded systems but becomes most apparent in confinement.
Despite a general good understanding of the behaviour in confinement, there still remain some gaps and some room for a further research. In our view, several stray questions which still await answers  are as follows :
\begin{itemize}
\item Our presentation was focused entirely on the situations in which the walls can be considered as geometrically perfect. In reality, of course, the walls can be geometrically rough or contain some contaminants acting as temporal traps for the environment and for the tracer particle.
\item The cross section of the channel may not be constant but be periodically varying along the channel. It is particularly interesting to analyse the case when this variation is sufficiently strong such that a channel consists of rather broad chambers separated by narrow passage tunnels. In this case, one expects interesting effects to emerge due to the hindered passages of the tracer particle through the bottlenecks (see, e.g., Ref. \cite{rubi} for more details).
\item An interesting and experimentally-relevant generalisation of the analysis presented here concerns the situations when the tracer particle is not subject to a constant external force but is active, being either a molecular motor which carries a cargo or is self-propelling itself in a crowded environment (see, e.g., Ref. \cite{dev} and references therein).
\item   Dynamics in ramified narrow channels still remains rather poorly understood. In particular, it is not very clear what will happen in situations when a biased tracer particle moves in a quiescent environment evolving in branching narrow channels.
\item The numerical analysis in Ref. \cite{kuster} of the case when several biased tracer particles move randomly in a narrow crowded channel seems to focus exclusively on the situations with a big driving force. On the other hand, there are all reasons to believe that self-clogging will not emerge at lower forces. The question of specifying appropriate conditions for an effective functioning of such narrow channel devices with respect to the driven component is certainly important for many applications.  
\end{itemize} 

\section*{Acknowledgments}

The author thank Profs. P. Krapivsky and T. Ooshida for many helpful comments.

\end{document}